# Six Questions about 6G

Kimberley Parsons Trommler
Head of Thinknet 6G
Bayern Innovativ GmbH
trommler@bayern-innovativ.de

Matthias Hafner
Project Manager, Thinknet 6G
Bayern Innovativ GmbH
matthias.hafner@bayern-innovativ.de

Prof. Dr. Wolfgang Kellerer
Research Chair
Thinknet 6G
wolfgang.kellerer@tum.de

Peter Merz
Industry Chair
Thinknet 6G
peter.merz@nokia.com

Sigurd Schuster
Head of Working Group for Digital
Infrastructure and Services
MÜNCHNER KREIS
sigurd.schuster.ext@nokia.com

Josef Urban
Member of the Research Committee
MÜNCHNER KREIS
josef.urban@nokia-bell-labs.com

Uwe Bäder
Advisor to the Workshop Committee
Uwe.Baeder@rohde-schwarz.com

Dr. Bertram Gunzelmann
Advisor to the Workshop Committee
bgunzelmann@apple.com

Andreas Kornbichler
Advisor to the Workshop Committee
andreas.kornbichler@siemens.com

*Although 5G ("Fifth Generation") mobile technology is still in the rollout phase, research and development of 6G ("Sixth Generation") wireless have already begun. This paper is an introduction to 6G wireless networks, covering the main drivers for 6G, some of the expected use cases, some of the technical challenges in 6G, example areas that will require research and new technologies, the expected timeline for 6G development and rollout, and a list of some important 6G initiatives world-wide. It was compiled as part of a series of workshops about 6G held by Thinknet 6G [1] and MÜNCHNER KREIS [2] in 2021.*

## I. What are the main drivers for 6G?

The current generation of mobile communication, 5G, is the first to move the focus away from individual end-user communications, placing more emphasis on industrial applications such as advanced manufacturing (Industry 4.0), logistics, transportation and e-Health. 6G will expand the capacity and speed of the networks further to enable applications with significantly higher networking requirements, such as real-time digital twins, full autonomous driving, or personal Body Area Networks (BAN). 6G moves the focus from machines to human beings and to their interaction with the environment around them, by supporting highly available, reliable, and secure communication with a dynamic, intelligent infrastructure.

6G will support societal goals and the UN Sustainable Development Goals (SDGs), such as environmental sustainability, efficient delivery of healthcare and education, reduction in poverty, hunger and inequality and, in particular, SDG #9 to "build resilient infrastructure, promote inclusive and sustainable industrialization and foster innovation".

As the Hexa-X project details [3], there is a clear and strong consensus among major 6G stakeholders that network technology must support and accelerate the transition to a better and more sustainable world, by ensuring that the following aspects are baked into the design from the beginning:

- Improved connectivity for a better and more sustainable world
- Built-in trustworthiness in an open society
- Digital inclusion that serves all populations
- Pervasive AI for human-centric and trustworthy automation and intelligence everywhere
- Mobile communications as a global ecosystem and success story

In addition, there are regional, national and international geopolitical aspects in the background. Technological sovereignty, for example, is an important issue for Germany and the rest of Europe. Acceptance and trust in the technology is important in all societies.

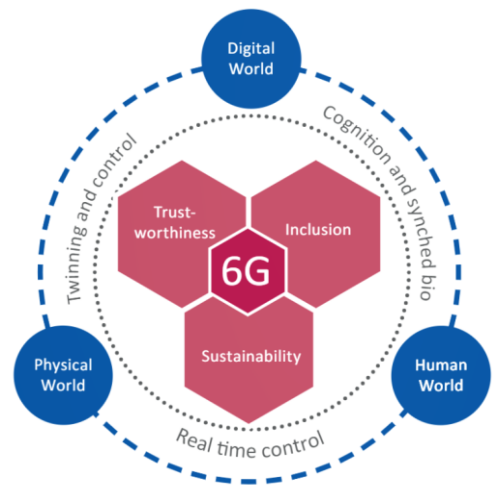

Figure 1: A vision of 6G
Source: Hexa-X Project [3]

## II. WHAT ARE THE EXPECTED USE CASES FOR 6G?

While 6G is the successor technology for 5G, it will be designed to meet requirements for use cases that allow creating and capturing value and that are currently not achievable with 5G. For example, Hexa-X listed those use cases and grouped them into 5 clusters [3] (see figure below):

- **Sustainable development:** 6G will be designed to address both environmental sustainability and the sustainable development of human societies, for example by helping vertical industries reduce their environmental impact or by enabling global digital inclusion. Use cases related to dematerialisation (e.g., e-Health, telepresence), more efficient use of resources (e.g., dynamic supply chains), optimization (e.g., flexible production), and monitoring of the environment contribute directly to environmental sustainability.

- **Massive twinning** is the extension of digital twin technology currently used in manufacturing to a wide variety of use cases, to provide a full digital representation of the environment, transportation, logistics, entertainment, social interactions, digital health, defense, public safety, smart cities, or food production. Digital twinning requires large data volumes, extreme performance, reliability and trustworthiness well beyond the current 5G capacities.

- **Immersive telepresence** is the ability to be present and interact with other people, anytime, anywhere, using all the senses (vision, hearing, touch, smell, taste, and proprioception). It enables people to interact with each other, with the physical world and with the digital world, necessitating a seamless unification of the physical, digital and human worlds. Use cases such as holographic presence or immersive reality place extreme bandwidth and latency requirements on the new network.

- **Cobots** are the next logical extension of robots, machines that form symbiotic relationships amongst each other or that better serve the needs and demands of humans in day-to-day interactions. By collaborating either with other robots or with humans, cobots can fulfil tasks more sustainably and more intelligently. These use cases extend 5G into scenarios with an increased number of devices, increased latency requirements, sophisticated coordination demand, and high requirements on trustworthiness.

- **Local trust zones** for human & machine: Some of the new 6G use cases involve sensitive information that requires better protection than classical IT security infrastructures can provide. 6G will need to provide fine-grained "trust zones" to protect sensitive data while still permitting valid use of that data. For example, individualized medical data from sensors can be used for personalized health monitoring, diagnosis, and therapy. But the network and applications must protect this data from misuse while still providing access to the data under pre-defined rules.

- **Evolving use cases** that are not yet known today. If you build it, they will come. As both the 6G network and other new technologies develop, new use cases that we currently can't envision are sure to arise. The network will need to be flexible enough and extendable enough to enable these new use cases as they appear.

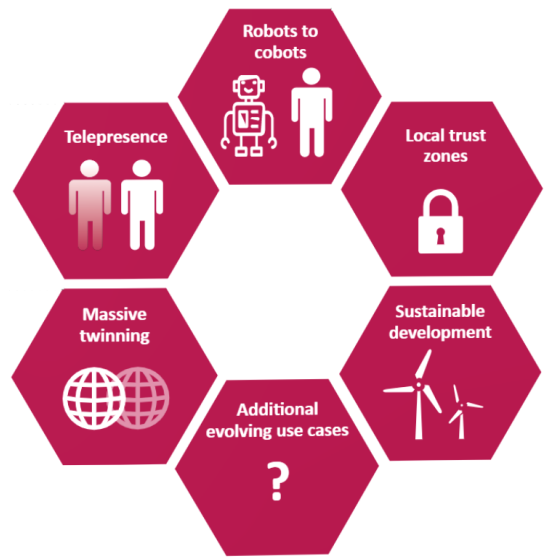

Figure 2: Use case families identified by Hexa-X
Source: Hexa-X Project [3]

## III. WHAT ARE THE MAIN TECHNICAL CHALLENGES IN 6G?

While the standards and new technologies for 6G are still in their infancy, current discussions around 6G include the following challenges and related research needs:

- The use of new (higher) frequency ranges: These new frequencies will require new antennas, and the use of higher frequencies will require more antennas than 5G, which complicates site planning and rollout.

- 6G will probably include not only more antennas than in 5G, but also a much higher number of end devices (e.g., IoT devices), leading to requirements for more dense coverage. The current network architecture may evolve to include peer-to-peer connections in addition to the current end-device-to-network connections.

- Network functionality (already being virtualized in 5G) will spread over multiple cloud and edge infra-structures. AI will be used to optimize and dynamically reconfigure the network on-the-fly.

- The combination of virtualization, programmable networks, edge computing, and simultaneous support for public and private networks requires more complex data center planning and orchestration.

- The vast amounts of data expected from increased sensing, imaging and localization will need to be managed, stored, processed and exchanged between multiple heterogenous edge, cloud and network operators, while maintaining data security.

- Closer integration with other wireless network technologies such as WLAN and non-terrestrial networks (e.g., near-Earth orbit satellites) is envisioned.

- End-to-end security, privacy, trust and resilience over heterogenous network and cloud infrastructures and across multiple administrative domains need to be assured.

- 6G is expected to enable sustainability in other industry sectors and needs to limit its own environmental impact. In addition to energy consumption and the use of renewable energy, aspects such as e-waste, sourcing of raw materials, water consumption for production of devices and network equipment, and for the cooling of data centers need to be considered.

IV. WHAT ARE THE MAIN RESEARCH AREAS FOR 6G?

Meeting the above technical challenges requires fundamental research in multiple areas including radio technologies, network architectures, distributed cloud infrastructure, AI/ML, and security and trust concepts. We have identified research domains that are especially relevant for enabling the envisioned new use cases and capturing value by addressing societal and global challenges and economic growth opportunities. These domains include:

- The network as a multi-sensor
- AI/ML-native communication and network adaptation
- Security, privacy, trust, and resilience
- Sustainability and environmental impact.

*A. The network as a multi-sensor*

6G is expected to perform various sensing tasks and support high-precision localization with centimeter-level accuracy. This advanced sensing will be achieved through:
- extended radio capabilities that provide high-precision positioning and object detection
- correlating data available in the network such as traffic flows and patterns
- the massive number of sensors connected to the network (including personal area networks and even bio-digital interfaces) and the ability to integrate and correlate data from these sensors into a comprehensive overview
- combining and enriching sensing with data from network external sources such as weather services

Radio signals, specifically radar signals, and the reflection of those signals are widely used to "see" objects, for navigation and for ranging. Imagine now extending those radio sensing capabilities beyond specialized applications into the entire wireless communication infrastructure. Imagine the possibilities when every cell and every node in the network is capable of sensing the location of objects, their speed, and even their material composition.

The network could sense the position of other cars on the road, as well as pedestrians, and identify their position, speed and current direction, and could then issue warnings if any of these are about to collide. This data from the communication network could be augmented with data from the car's onboard cameras and sensors to provide improved "vision" even when it's dark or foggy, to see further ahead than the cameras, or to see around the corner. And this data from a single vehicle can be fused together with data from other vehicles, from congestion monitoring and from weather forecasts to provide a comprehensive view of the current traffic situation.

*B. AI/ML-native communication and network adaptation*

As part of the move from 5G to 6G, the use of AI will move from being an enhancement to being a fundamental element of the network design and optimization:
- An AI-based air interface could learn new waveforms for different frequencies on-the-fly, rather than having to go through the traditional process of algorithm design and implementation. Similarly, it may be possible for the network to learn new protocols, or to use AI/ ML to create new, better protocols during operation
- Across the entire network, AI/ML will be used for comprehensive end-to-end network automation and orchestration, including aspects such as AI-based reconfiguration on-the-fly without requiring human intervention
- At the application layer, positioning (as discussed above under "The Network as a Multi-Sensor") could be further improved based on machine learning

As a more concrete example, consider how one cell tower provides coverage for a certain area. The telecom company initially configures the antennas on that tower to cover the area as completely as possible. Over time, as traffic patterns change, the company could use AI to dynamically reconfigure the antennas, to optimize coverage on-the-fly, without requiring human intervention.

*C. Security, privacy, trust, and resilience*

Since the 6G network will blur the line between the human, physical and digital worlds, a breach in security could lead to a loss of information, loss of control over your devices, loss of money, loss of property, or even physical danger to people. In a worst-case scenario, cyber warfare could wreak havoc in the physical world, with a direct effect on national security. To address these concerns, the level of trust, privacy, security and resilience in 6G networks must be significantly higher than current state-of-the-art in today's data networks, which necessitates new development in many different areas:

- Trusted networking, such as trust management across multiple domains, standardization of trust and privacy models, and reliability across heterogenous networks/ technologies/providers
- Software-defined and AI-defined security and networking
- Security at the physical layer, such as distributed, cooperative security protocols or how to detect and mitigate against jamming
- How to empower human users to identify the data they're willing to share, and with whom, in an understandable manner

Consider one of the most important uses cases for 6G: human-centered mobile networking, with Body Area Networks consisting of on-body ("wearable") and in-body ("implant") devices. These devices are not only passive measuring devices (e.g., for monitoring blood pressure) but also active devices providing medication and treatment (e.g., insulin pumps or pacemakers), where failure or malicious intent can cause serious harm. Clearly, security for these Body Area Networks is critical. The network must provide communication to these devices for valid users/applications while blocking access to all others, with zero tolerance for errors.

*D. Sustainability and environmental impact*

Digital infrastructures, including 6G, influence sustainability and environmental impact via two main mechanisms. On the positive side (handprint), the 6G network provides the communication infrastructure that underlies digital applications to improve sustainability in other sectors, such as smart buildings, smart mobility, and smart cities. However, the production, operation, and decommissioning of end devices, networks and data centers use rare metals, and generate electronic waste and Greenhouse Gas (GHG) emissions. Some studies [4] estimate that the telecommunications industry was responsible for 2,3% of the global GHG emissions in 2020.

The energy consumption "per bit" was reduced from 4G to 5G, and we expect additional energy reductions per bit from 5G to 6G. This will help compensate the increased energy demand due to increased traffic in 6G, helping to flatten the curve.

As such, research and new technologies in the following areas are required:
- How to use the 6G network as a sensor to collect data for other sectors, to enable sustainability applications in other sectors (e.g., to monitor reforestation efforts)
- How to support new or extended use cases for other sectors (e.g., AR/VR, manufacturing) to enable them to reduce their GHG emissions through e.g., automation or optimization
- How to reduce e-waste, the use of raw materials, and water consumption for 6G devices and infrastructure
- How to reduce total energy consumption in the 6G network despite growing demand for connectivity and bandwidth, and induction and rebound effects
- Standards, policy, and regulatory measures that should be considered for a sustainable 6G infrastructure

A comprehensive framework to measure, monitor, and assess – in real-time – the environmental impact of 6G will be a prerequisite for many measures and techniques that rely on such data, such as AI-based energy management for 6G and the optimized use of power-saving features. Improving both the hand- and footprint of 6G will involve multiple simultaneous activities, including the use of renewable energy, reusing the heat produced by data centers, and a circular-economy approach for the materials and devices.

## V. WHAT'S THE TIMELINE FOR 6G?

Each new generation of cellular technology takes approximately ten years to develop, implement and roll out. While the 5G deployments are still underway, research for 6G has already begun. Since 6G is expected to hit the market around 2030, companies and research institutions need to begin their work on 6G without delay. The research and development of 6G technology, such as new hardware for the new frequency spectrums, is expected to require several years. The international standardization process then requires roughly an additional 2-3 years, with product development and testing in parallel and at the end of the 10-year cycle.

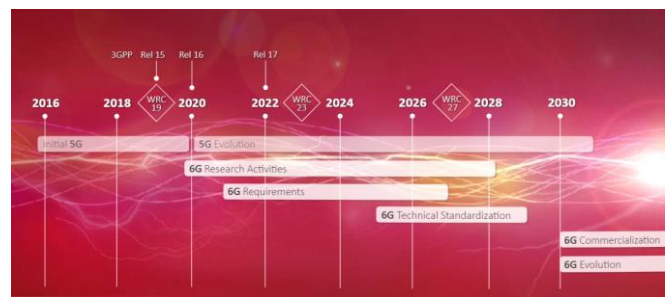

Figure 3: Expected timeline for 6G

The main milestones for 6G are expected to be:
- Initial 6G standards around 2025, although there will probably be a series of standards published over several years
- Frequencies/spectrum for 6G to be discussed at ITU WRC (World Radio Communication) conference in 2023, and then finalized/approved at the WRC in 2027
- Initial rollout beginning around 2030

### VI. WHAT'S THE CURRENT STATUS OF 6G IN BAVARIA, GERMANY, EUROPE AND THE REST OF THE WORLD?

6G activities and programs are ramping up across the world. Some of the more important initiatives (as of January 2022) include:

**Bavaria**:
The Bavarian Ministry of Economic Affairs, Regional Development and Energy announced a 6G initiative as part of the Hightech Agenda Plus, consisting of three pillars:
- The 6G Future Lab at the Technical University of Munich
- Thinknet 6G at Bayern Innovativ
- Funding calls for 6G research and development projects, for example the "Kommunikationsnetze der Zukunft" call for proposals in Spring 2021.

**Germany**:
- The German Federal government announced funding of 685 million Euro for 6G Research.
- The German Federal Ministry of Education and Research (BMBF) established four national 6G research hubs in August 2021, with funding of 200 million Euros. A national 6G platform is in the works, and a call for industry-led 6G projects was issued in fall 2021.
- A group of five Fraunhofer Institutes announced the 3-year, €8 million 6G Sentinel Project in February 2021.
- The Thinknet 6G Summit, the first 6G conference in Germany, was held in October 2021.

**Europe:**
- The EU's flagship 6G project, Hexa-X, officially kicked off in January 2021.
- The EU has granted over €95 million in funding to at least 20 projects involving over 156 organizations. 6GWorld's interactive graph [5] gives a good overview of these projects.
- The 6G Smart Networks and Services Industry Association (6G-IA) announced the start of the Smart Networks and Services Joint Undertaking (SNS JU) and the adoption of the Smart Networks and Services Research and Innovation Work Programme 2021 – 2022 (SNS R&I WP 2021-2022) in December 2021.
- The EU Smart Networks and Services Phase 1 call opened in January 2022.
- The 6G Flagship at the University of Oulu, Finland and the 6G Innovation Centre at the University of Surrey, England support collaboration between research and industry for 6G.

**APAC:**
- China launched an experimental satellite for 6G in November 2020.
- The China National Intellectual Property Administration (CNIPA) made an announcement on World Intellectual Property Day (April 26[th], 2021) that they are the leader in the 6G race. Of the 38000 patents related to 6G technologies, 35% are from Chinese companies.
- Korea's Ministry of Science and ICT announced a 6G R&D Implementation Plan with ca. 142 million Euro funding, with pilot projects planned for 2026.
- India's Department of Telecommunications established a 6G Technology Innovation Group with members from the government, academia, industry associations and the Telecom Standards Development Society of India.

**USA:**
- The Next G Alliance, an ATIS (Alliance for Telecommunications Industry Solutions) initiative, aims to establish North American preeminence in 5G evolution and 6G development.
- The US National Science Foundation announced the Resilient & Intelligent NextG Systems (RINGS) program to fund research in areas with significant impact on emerging next generation wireless and mobile communication.

**International:**
- The USA and Korea agreed to encourage joint R&D on 6G.
- The USA and Japan announced a $4.5 billion partnership for 5G and 6G networks.
- The NGMN Alliance (Next Generation Mobile Networks Alliance) published a paper about 6G drivers and vision and has established 6G working groups.

ACKNOWLEDGMENTS

Thinknet 6G at Bayern Innovativ is funded by the Bavarian Ministry of Economic Affairs, Regional Development and Energy.

Thinknet 6G and MÜNCHNER KREIS thank the members of the workshop committee for sharing their expertise and for their contributions to this paper.